# Abiotic streamers in a microfluidic system

Nandini Debnath[a], Mahtab Hassanpourfard[b], Ranajay Ghosh[c], Japan Trivedi[d], Thomas Thundat[b] and Aloke Kumar[ae]*

[a]Department of Mechanical Engineering, University of Alberta, Edmonton AB T6G 2G8, Canada

[b]Department of Chemical and Materials Engineering, University of Alberta, Edmonton AB T6G 2G8, Canada

[c]Department of Mechanical and Aerospace Engineering, University of Central Florida, Orlando FL 32816, USA

[d]Department of Civil and Environmental Engineering, University of Alberta, Edmonton AB T6G 1H9, Canada

[e]Department of Earth and Environmental Engineering, Columbia University, New York NY 10027, USA

**<u>*Corresponding author email:</u>** aloke.kumar@ualberta.ca

**Abstract:** In this work, we report the phenomenon of formation of particle aggregates in the form of thin slender strings when a polyacrylamide (PAM) solution, laden with polystyrene (PS) particles is introduced into a microfluidic device containing an array of micropillars. PAM and dilute solution of PS beads is introduced into the microfluidic channel through two separate inlets and localized particle aggregation is found to occur under certain conditions. The particle aggregates initially have a string-like morphology that remain tethered at their ends to the micropillar walls, while the rest of the structure remains suspended in the fluid medium. It is this morphology that inspired us to name these structures streamers. The flow regime under which streamer formation is observed is quantified using through a phase diagram. We discuss the streamer formation time-scales and also show that streamer formation is likely the result of flocculation of the PS beads. These streamers can serve as excellent model systems to study a biological phenomenon by the same name.

# Introduction:

The study of flow induced structure formations in colloidal dispersions, either biological or non-biological, is a topic of substantial contemporary interest[1-3]. Entities called bacterial streamers belong to the first category and represent an extremely complex and intriguing biological phenomenon resulting from an interplay between life processes, fluid-structure interactions and material nonlinearities[4-8]. Structurally, bacterial streamers are slender filamentous aggregates primarily comprising of bacterial cells encased in matrix of self-secreted extra-cellular polymeric substances (EPS)[4] and formed typically under sustained hydrodynamic flows (both turbulent and low Reynolds number ($Re$ <1) conditions)[1, 9]. Formation of these biological streamers in low Reynolds number flows ($Re$ <1) is a topic of contemporary interest as they can have significant impact on the performance of filtration units[7, 10] and biomedical devices[1, 6]. The overall complexity of the material behavior of bacterial streamers arises due to the nature of EPS, which is a biological soft material with highly nonlinear stress-strain relationships[11-14], and the embedded live bacterial cells forming a multiphase composite soft media.

Despite significant work from various research groups[5, 6, 11, 15], formation of bacterial streamers remains an unresolved issue. On the non-biological front, industrial operations on particles dispersed in rheologically complex fluids have motivated the study of alignment and aggregation of abiotic beads in viscoelastic media[16-18]. The aqueous solution of polyacrylamide (PAM) is an excellent example of a complex or viscoelastic liquid and the hydrodynamics of such fluids have been an area of great importance both from fundamental and applied science perspectives[3, 19, 20]. An interesting feature of particle laden flows of PAM or other viscoelastic liquids vis-à-vis Newtonian liquids is the observation of particle flow-induced alignment or pattern formation even in the dilute regime. Vermant and Solomon[3] discuss some such studies. For example, Belzung et al. observed a butterfly pattern formation at certain shear rates when latex particles were added to PAM[21]. Later, Hoekstra et al. examined the reversible aggregation in a viscoelastic flow induced two-dimensional particulate network[2]. However, addition of particles to complex fluid can considerably enhance the hydrodynamic complexity and rheological properties of the system. For instance, Umeya and Otsubo found that when a certain critical shear rate is exceeded, the apparent viscosity increases due to an irreversible adsorption of PAM molecules by particles[22]. Other than the viscoelastic nature of the suspending media, and particles, the structure of the flow channel can itself introduce complicated flow patterns[23]. This structural effect of the confining geometry could be of particular interest in microfluidic devices with parallel micro-pillars which serve as analogs to diverse microporous media[24, 25]. Such systems are already well known to be vulnerable to a streamer mode of fouling and clogging as described earlier[25]. In this paper, we study the effect of this type of device geometry on a synthetic flow system, which is topically similar to the bacteria floc laden flow in terms of mechanical constitution. This helps us to test the hypothesis that physical factors are primarily responsible for formation of the unique slender structures associated with biological streamers, and also post-formation dynamics such as clogging.

In this work, we investigated the combined flow of PAM (0.2% w/w) and polystyrene (PS) bead solutions (0.8 % w/w) through a microfluidic device containing an array of micropillars. The relaxation time scale of the PAM solution will be denoted by $\lambda$. The two solutions were introduced into the microchannel through separate inlets (PAM at volume flow rate $Q_{PAM}$ and PS bead solutions are $Q_{PS}$) and the flow through the device is characterized by a very low Reynolds number ($Re<<1$). It was found that flow regimes characterized by $Q_{PAM} \leq$ $Q_{PS}$, the polystyrene nanoparticles aggregated to form thin slender strings, which we have named as abiotic streamers. Two distinct time-scales are observed in the streamer formation process. The streamer formation time-scale is of the order of a few seconds. At much larger time-scales of approximately an hour, the initial slender structures mature to clog the pores of the device. It was determined that streamer formation was the result of flocculation (irreversible clustering) of PS beads analogous to biological flocculation[5] and abiotic streamer formation correlated well with the extent of PS bead flocculation. Discovery of this phenomenon of abiotic streamer formation verified the hypothesis that physical factors dominate the formation of bacterial streamers. These abiotic streamers can also serve as model systems to understand bacterial streamers as well as point towards a novel route to clogging in synthetic systems like these.

## Materials and Methods:

*PAM and PS solution preparation*

Polymer solution was prepared by dissolving 1g of polyacrylamide powder (PAM: A-8354, Kemira, AB, Canada) into 500mL of normal tap water. Then, the PAM solution (0.2% w/w) was agitated at 600 rpm with an overhead stirrer (Caframo, ON, Canada) for about three hours to ensure proper mixing (Fig. S1). Further, Fluorescein sodium salt (Excitation at 460nm & Emission at 515nm) (Sigma-Aldrich, ON, Canada) was also added to the PAM solution to make it green fluorescent. The particle solution was prepared by diluting 200nm amine-coated PS beads (Excitation at 580nm, Emission at 605nm) (Life Technologies, ON, Canada) with deionized water till a 0.8% w/w suspension of PS beads was attained. Under fluorescence microscopy PAM appeared green, while the particles appeared red. PAM and PS solutions were separately injected into the two inlets of microfluidic channel by using a dual-syringe pump (Harvard Apparatus, MA, USA). The flow rate was controlled by the pump to maintain a constant volume flow rate for both syringes. The syringes (3 mL each) were connected to the microchip by Tygon tubing (ID 0.01mm, Fisher Scientific, AB, Canada).

*Microfluidic chip fabrication*

The polydimethylsiloxane (PDMS, Sylgard 184, Dow Corning, NY, USA) chip with the required microchannel design was prepared by conventional photolithography from a 4″ silicon master mold. Our microfluidic device contained two inlets (Fig 1a) and a central channel containing an array of micropillars with an out-of-plane height *h* and diameter *d* of 50μm each (Fig. 1b,c). The distance between center of pillars and gap between two consecutive rows, *s*, was 75μm. Width, *w*, of the microchannel was 625μm. The array consisted of 8 (across breadth) X 50 (across length) micropillars (Fig. 1b,c). The PDMS stamps and cover slip were bonded together by using oxygen plasma-activated bonding for 30 seconds. Further, they were annealed at $70^0$ C to ensure proper sealing as described by Hassanpourfard et al.[5].

*Microscopy*

The entire microchip was placed on a stage of an inverted optical (Nikon Eclipse Ti) microscope and a confocal microscope (Olympus IX83) (Fig. 1d). Fluorescent microscopy directly probe imaging and videography by using either a GFP Long-pass Green filter cube or Texas Red filter cube (Nikon & Olympus).

*Ex-situ rheological measurements*

A standard parallel-plate rheometer (C-VOR 150 Peltier Bohlin Rheometer, Malvern instruments, USA) was used for the viscoelastic measurements to investigate the rheological effects for the polymer-mediated streamer. To mimic our microfluidic experiment 10mL sample of PAM (0.2%) and PS (0.8%) were vortexed together to ensure homogenisation before the measurement. All experiments were performed at room temperature ($20^0$ C).

## Results:

Our microfluidic device had two inlets. PAM was injected from the first port at a (volumetric) flow rate of $Q_{PAM}$ and amine coated PS bead suspension in deionized (DI) water from the second port at a flow rate of $Q_{PS}$ (Fig. 1). The two streams merged at a Y-shaped intersection (see Supplementary Video 1) before entering the central section of the microfluidic channel (Fig. 1a). The central section of the device consisted of an array of PDMS micropillars in a staggered grid pattern. The micropillars had a diameter (*d*) of 50 microns and were spaced 75 microns apart (*s*) (Fig. 1b,c). The average velocity scale ($\bar{U}$) in the device is defined by the relationship $\bar{U} = (Q_{PAM} + Q_{PS})/(w \times h)$, and the flow rates $Q_{PAM}$ and $Q_{PS}$ were maintained at a level such that the resultant flow in the device was in the creeping flow regime (*Re*<<1). As $Q_{PAM}$ and $Q_{PS}$ are the native experimental variables, all results will be stated in their terms. The average velocity scale ($\bar{U}$) will only be used for discussion purposes. The relaxation time scale, $\lambda$, for PAM was determined through rheological studies and found to be 13s (see Fig. S2). This device configuration was earlier used by Hasanpourfard et al.[5], where they showed that bacterial flocs flowing through the device could adhere to the micropillar walls and get rapidly sheared by the background flow to form bacterial streamers. These biological streamers are slender filamentous structures with a high aspect ratio (length/radius ~ *O*(10)), where bacteria are encased in EPS (Fig 2a). Our experiments reveal that under certain conditions the flow of PAM and PS in our device could also lead to the formation of streamers, which are morphologically similar to biological streamers.

When PAM (0.2% w/w) and PS beads (0.8% w/w) solutions flow through the microfluidic device at flow rates of 10μL/h and 40μL/h respectively, we observed that there was an immediate localization of PS beads near micro-pillar walls in the form of string-like filamentous structures (Fig 2b). The structures are tethered to the micropillar walls, while the rest of the structure is immersed in the surrounding fluid. Such particle aggregation is also irreversible as it was observed that if the flow stopped, the structure itself remained undisrupted. For the streamers shown in Figure 2b, the particular dilutions for PAM (0.2% w/w) and PS (0.8% w/w) were carefully chosen through a series of optimization experiments (data not shown) that resulted in stable and observable streamer formation. These experiments also revealed that streamers were not observed below a certain cutoff PS concentration. However, when they successful form, these structures tend to be pervasive throughout the microfluidic device i.e. they occurred between a majority of the first 10 rows of micropillars. Figure 2,b,c compares abiotic streamers with their bacterial counterparts and demonstrates the morphological similarities between the two. We have already extensively reported on the formation of bacterial streamers[1, 5, 7, 25]. Structurally, bacterial streamers are composed of bacteria (Young's modulus ~ *O*(100 MPa)[26] encased in extra-cellular polymeric substances (EPS) (Storage modulus ~ *O*(1Pa)[27] . Thus, in abiotic streamers PS beads (Young's modulus ~ *O*(100MPa)[28] can be considered as surrogates for bacterial cells, and PAM (Storage modulus ~ $O$(1Pa), see Fig. S2) can be considered as a surrogate for EPS.

Two distinct time-scales were observed for the abiotic streamer under present experimental conditions (Fig. 3) (also see Supplementary Video 1): (i) Streamer formation time-scale ($\tau_{form}$), which was of the order of the a few seconds i.e. $\tau_{form} \sim O(10^0 s)$. At these short time-scales the streamers have a high aspect ratio (length/radius ~ $O$(10)). (ii) Clogging time-scales ($\tau_{clog}$), which was of the order of several minutes i.e. $\tau_{clog} \sim O(10^3 s)$. As the initial structures mature by mass accretion the slenderness ratio of the streamers decrease and after approximately 60 minutes streamers engulf the entire pore-space of the device and the length to radius ratio approaches unity. This progression in streamer structure is similar to those displayed by bacterial streamers[5]. Figure 4a shows a two-color superimposed confocal image of the streamer after approximately 60 mins of experimentation. The fluorescein mixed PAM appears green, while the PS beads appear red. It can be clearly seen that the PS beads localize around the pillars, but PAM solution is well mixed everywhere (Fig 4b,c). These experiments were complemented by two control experiments. In the first case only PAM solution was flown through the device and in the second only PS solution was flown through the device. In neither case streamer formation was observed, thus confirming that the combined flow of both PAM and PS was required for streamer formation (Fig. S3).

Our experimental setup was designed to allow us to vary $Q_{PAM}$ and $Q_{PS}$ independently of each other. To evaluate the role of the volumetric flow rates on streamer formation, experiments were conducted by varying the two quantities while keeping the PS bead and PAM concentrations fixed at 0.8% and 0.2% respectively. This constant composition flow under the current setup allows us to develop a formation plot phase diagram. Figure 5 represents this formation plot, where a red triangle and blue circle are used as binary markers for streamer formation and its lack of respectively. This plot shows an intrinsic asymmetry of streamer formation with respect to flow rates. Streamers form only when $Q_{PS} > Q_{PAM}$. These streamers are remarkably stable and eventually form mature streamers. On the contrary, the system is marked by absence of streamers when $Q_{PS} \leq Q_{PAM}$, even after a maximum observation time-scale of 60 minutes. Quite interestingly, at the phase boundary, i.e. when $Q_{PS} = Q_{PAM}$ streamer formation was not observed. Another notable feature of these experiments was its repeatability (see Fig. S4 for repeatability data). Streamer formation appeared to be a strong function of the surface chemistry of the PS beads. When we replaced the amine-coated 200nm PS beads with carboxylate-coated 500nm PS beads, streamer formation was not observed at any flow-rate.

Observing the striking similarity of abiotic streamers for the amine-coated PS beads with their bacterial counterparts (Fig. 2b,c), particularly the striking similarity in the time-scales observed by Mahtab et al.[5], we hypothesized that flocculation of the PS beads had a role to play in this phenomena. Two competing phenomena exist which control the deposition of PAM on PS beads – diffusion and convection. In order to ascertain the extent of one over other, we compute the non-dimensionless Peclet number (*Pe*) for the current system which determines the deposition mechanism of the more mobile PAM polymer on the PS bead. To this end, *Pe* is the ratio of two time scales

$$Pe = \frac{\tau_{diff}}{\tau_{conv}} = \frac{\left(\frac{4R^2}{D}\right)}{\frac{R}{\bar{U}}} = \frac{4R\bar{U}}{D} = \frac{4 \times 200 \times 10^{-9} \times 500 \times 10^{-6}}{10^{-10}} = 40 \times \frac{10^{-11}}{10^{-10}} \sim O(1) \qquad (1)$$

where *R* is the radius of the bead and *D* is diffusivity of PAM in water. This shows that the deposition is largely diffusion dominated. We simulate this diffusion dominated deposition conditions for floc formation by mixing amine coated PS beads (0.8% w/w) in PAM solution (0.2% w/w) with quiescent steady state conditions lasting about a few minutes. The solution was then observed under optical microscope. The resulting solution showed profuse flocculation thus verifying our hypothesis. The distributions of floc size are presented in Figure 6(a-c) with accompanying optical images. These experiments were repeated for three different volume ratios of PAM and PS beads – PAM:PS=1:4,1:1 and 4:1. These experiments showed that higher relative concentration of PS beads favor larger flocs (Fig. 6a). The average (mode) diameters of flocs for these concentrations were respectively 16.35µm, 4.46µm and 1.08µm. However, note that although diffusion is the primary transport process, which brings the PAM to the PS bead surface, the ultimate adhesion would be determined by the adsorption kinetics of the PS surface and would thus be affected by the chemical treatment of the bead. In order to contrast this effect, we repeated the same experiments with carboxylate-coated PS beads. This time, the system exhibited a complete absence of flocculation even after several hours of observation.

Under the microfluidic flow conditions, the formation plot shows extensive streamer formation when $Q_{PS} > Q_{PAM}$ and only for amine coated PS beads. Since no streamers formed under any flow condition for carboxylate coated PS beads, and considering the floc shearing as the observed mode of streamer formation in our earlier experiments on bacterial systems, we conclude that flocculation is the primary mechanism of abiotic streamer formation as well. A simple qualitative explanation based on flocculation also explains the formation plot. First, assuming low *Re* just ahead of the Y-junction leads to a relatively well-defined interface between the PS and PAM flow with cross diffusion dominating the mixing process, Figure 1a. The dashed line (∂B) in Figure 1a depicts this interface, which partitions the total width of the channel (*w*) into the widths $w_{PS}$ and $w_{PAM}$. From interface continuity of tangential velocity at ∂B we have:

$$\frac{Q_{PS}}{w_{PS}} = \frac{Q_{PAM}}{w - w_{PS}}, w_{PS} = \frac{Q_{PS}}{Q_{PS} + Q_{PAM}} w \qquad (2)$$

Now it is clear that as PAM velocity is increased, the PS bead flow channel narrows down. Since concentration of PS beads is constant, this means the total number of beads at a given cross section decreases. Assuming low *Pe*, we directly deduce from our quiescent flow experiments that this would mean lesser number of PS beads per unit PAM and thus lesser probability of larger flocs which are the primary structural units to precipitate the floc mediated mechanism outlined earlier by the biological counterpart[5].

In conclusion, using a system of PS beads as a surrogate for bacteria and PAM as EPS we have demonstrated for the first time the universality of streamer formation and the consequent importance of physical factors governing the formation dynamics as well as overall geometry. First we observed that only amine coated PS beads showed streamer formation. This correlated with lack of flocculation in the carboxylate-coated system thereby pointing to flocculation as a primary mechanism of streamer formation. This makes flocculation a far more general mechanism of streamer formation than previously reported[5]. This flocculation hypothesis also provided a parsimonious explanation of the formation plot. The plot highlighted an inherent asymmetry in the formation with respect to the flow rate of the polymer (EPS surrogate) and particulate (bacterial surrogate) which was explained by a straightforward flocculation hypothesis and diffusion dominated mixing conditions.

# Figures:

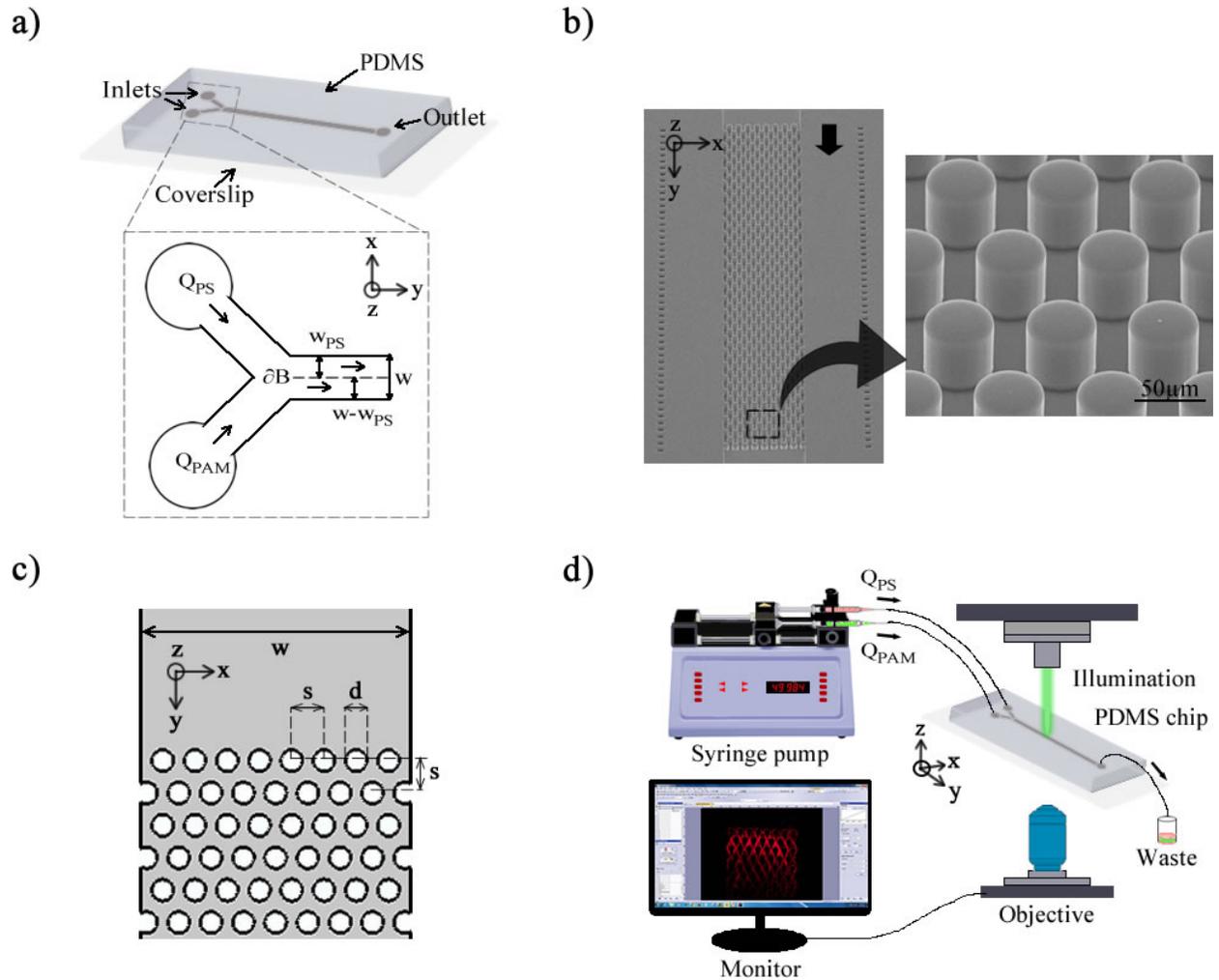

**Figure 1. a)** A schematic of the microfluidic device having two inlets and one outlet. The blow-out section shows the entry Y-channel where mixing process occurs. Polymer solution (PAM) and polystyrene solution (PS) were injected with a constant volume flow rate $Q_{PAM}$ and $Q_{PS}$. Total width of the channel is $w$ and is partitioned into a section with PAM ($w_{PAM}$) and PS beads ($w_{PS}$) **b)** SEM image of the microchannel clearly shows the uniformity of micropillars. **c)** Top-view of the microfluidic device. The diameter ($d$) and out of plane height ($h$) of the pillars both are 50μm and width ($w$) of the device is 625μm. The distance between the center of the pillars ($s$) and two rows of the consecutive pillars ($s$) is 75μm. **d)** A schematic of the total experimental set up. PAM (green) and PS (pink) solutions were transported through two inlets of the microfluidic device with a dual syringe pump which generates a constant volume flow rate $Q_{PAM}$ and $Q_{PS}$.

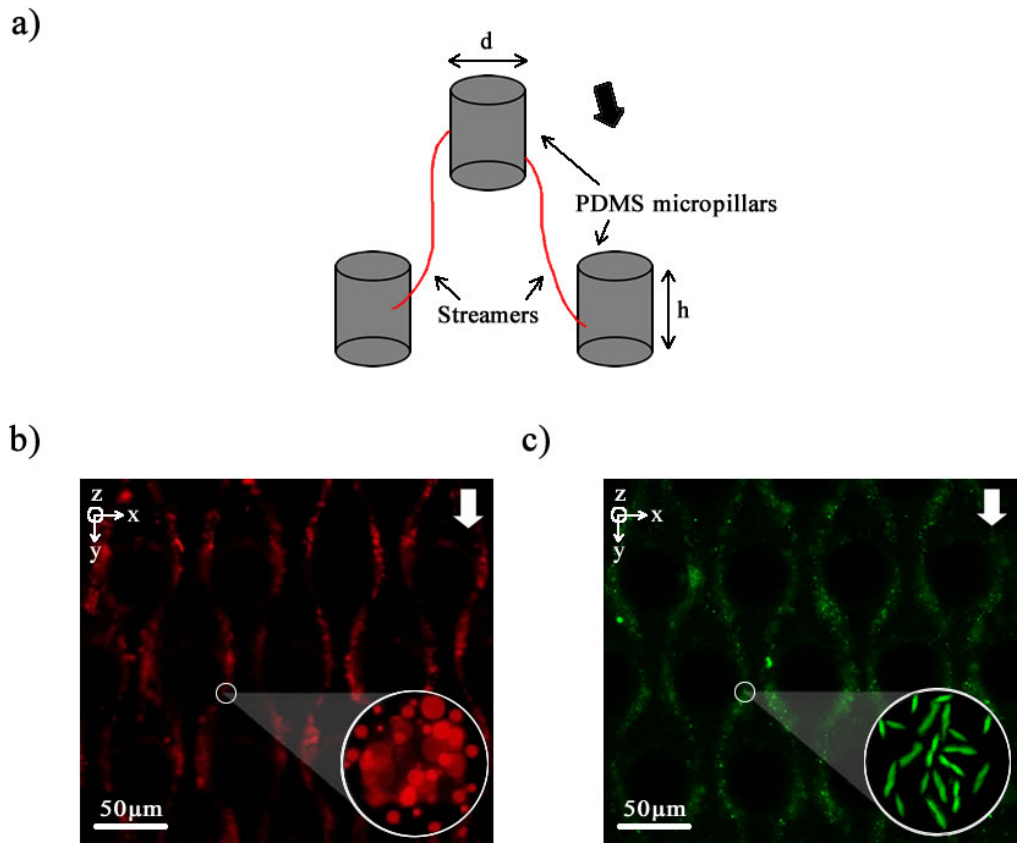

**Figure 2: a)** A schematic of streamers forming between micropillars. **b)** Abiotic streamer containing PAM (0.2%) and PS nanoparticles (0.8%). **c)** Bacterial floc mediated streamer containing EPS and bacterial cells.

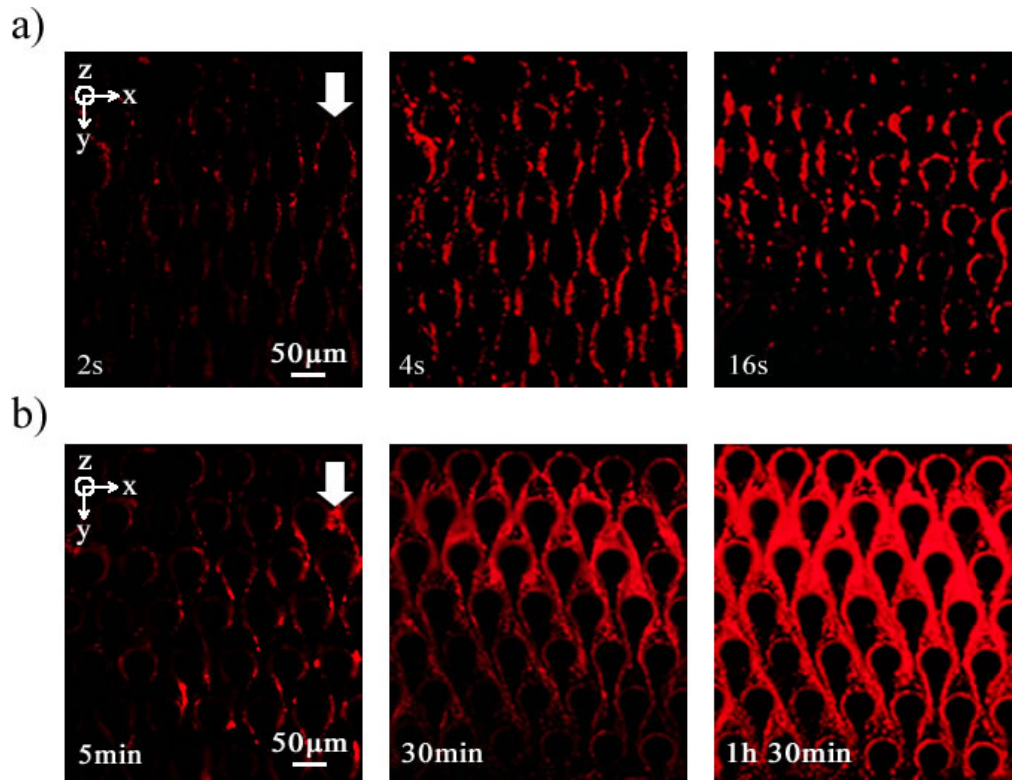

**Figure 3.** Streamer formation imaged using the Texas Red filter cube **a)** PS beads localize to form a streamer like structure. The time-stamp also indicates that these streamers form over a very short time-scale (a few seconds) for flow rate $Q_{PAM}$=10µL/h and $Q_{PS}$=40µL/h. **b)** Over larger time-scales almost 'solid-like' streamers are seen and they accumulate more PS beads over time and finally become very thick mature structures for the same flow rate.

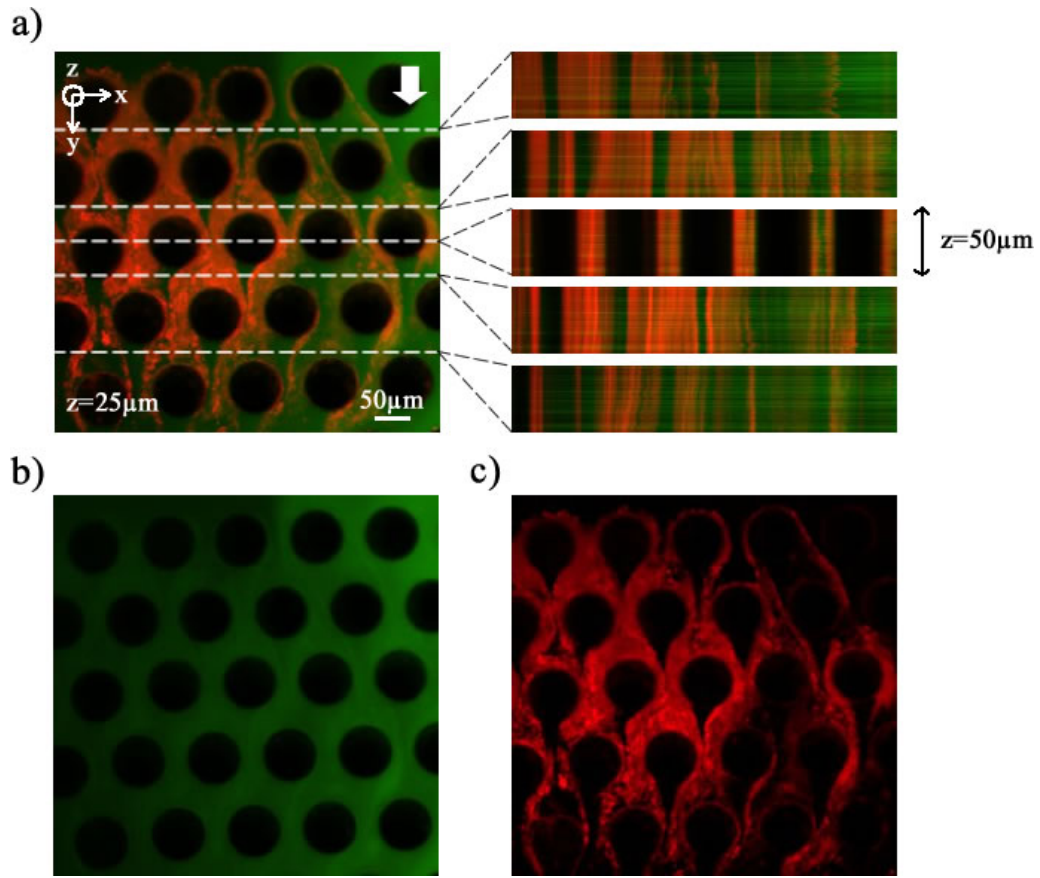

**Figure 4. a)** Two-color imaging confocal imaging of a streamer after 1 hour of experimentation. White-dashed line indicates the *y*-location at which the corresponding *z-x* plane is depicted. Image of the same location using **b)** FITC Green filter cube and **c)** Texas Red filter cube.

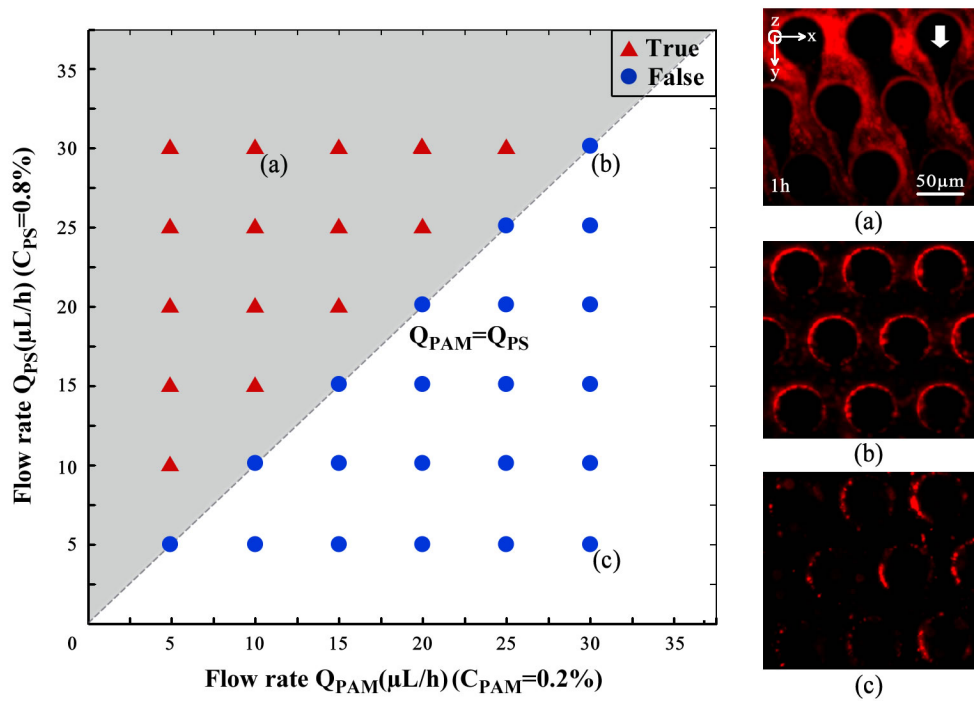

**Figure 5.** Formation phase diagram of streamer formation over different flow rates for $Q_{PAM}$ and $Q_{PS}$ when the concentration is fixed. Red triangle represents true streamer formation and blue circle indicates no streamer formation. The side-bars depict the optical microscope observations at the three located delineated in the formation plot: **a)** Streamer formation after one hour for $Q_{PAM}$=5μL/h, $Q_{PS}$=20μL/h; **b)** No streamer formation at phase boundary; **c)** No streamer formation after one hour for $Q_{PAM}$=30μL/h, $Q_{PS}$=25μL/h.

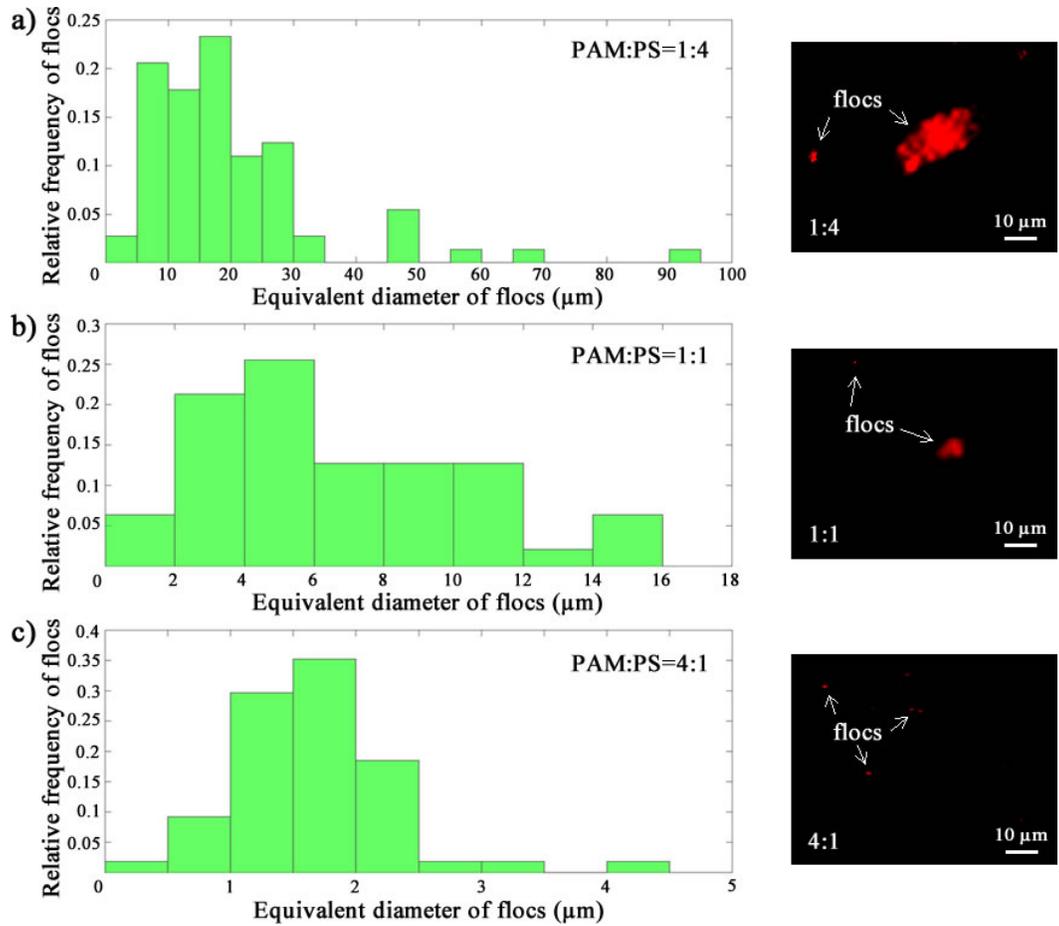

**Figure 6.** Relative frequency histogram of PS (0.8%) flocs and corresponding floc image after mixing with PAM (0.2%). **a)** PAM(0.2%): PS(0.8%)=1:4, the median and mode for this relative frequency histogram are 17.05µm and 16.35µm, (floc count 75) respectively and the streamer forms for flow rate $Q_{PAM}<Q_{PS}$. **b)** PAM(0.2%): PS(0.8%)=1:1, the median and mode are 5.67µm and 4.46µm (floc count 50), respectively and the streamer does not form by introducing same flow rate($Q_{PAM}= Q_{PS}$). **c)** PAM(0.2%): PS(0.8%)=4:1, the median and mode are 1.62µm and 1.08µm (floc count 54), respectively and the streamer does not form for any flow rate.

# Supplementary Information for

# Abiotic streamers in a microfluidic system

Nandini Debnath[a], Mahtab Hassanpourfard[b], Ranajay Ghosh[c], Japan Trivedi[d], Thomas Thundat[b] and Aloke Kumar[ae]*

[a]Department of Mechanical Engineering, University of Alberta, Edmonton AB T6G 2G8, Canada

[b]Department of Chemical and Materials Engineering, University of Alberta, Edmonton AB T6G 2G8, Canada

[c]Department of Mechanical and Aerospace Engineering, University of Central Florida, Orlando FL 032816, USA

[d]Department of Civil and Environmental Engineering, University of Alberta, Edmonton AB T6G 1H9, Canada

[e]Department of Earth and Environmental Engineering, Columbia University, New York NY 10027, USA

**Corresponding author email:** aloke.kumar@ualberta.ca

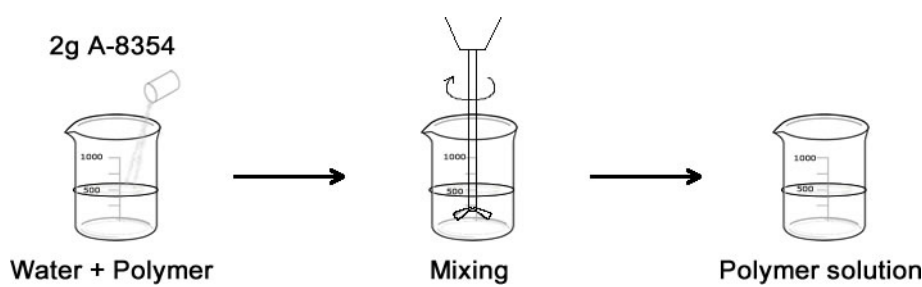

**Figure S1:** Schematic of the preparation of polymer solution at room temperature. 2g of polymer were mixed with 500ml of normal tap water (0.2% w/w PAM solution) and mixed at 600 rpm for at least 3 hours to ensure homogenization.

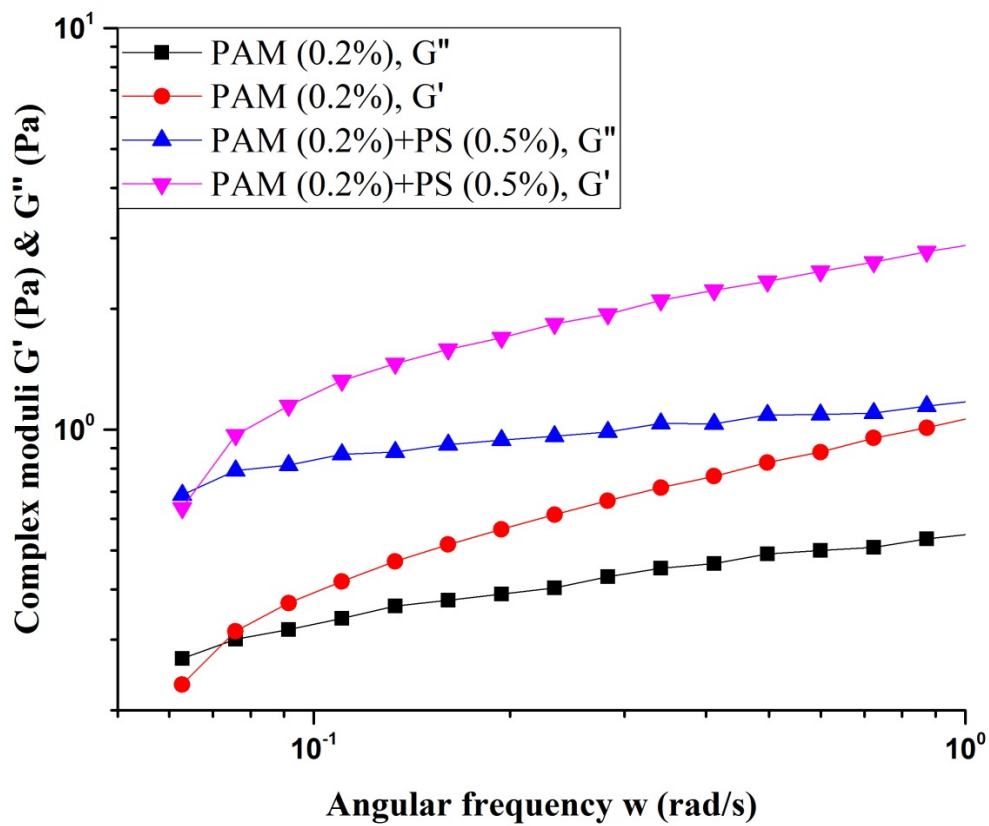

**Figure S2:** Comparison of the complex moduli for PAM (0.2%) and PS (0.5%) solution with PAM (0.2%) solution as a function of angular frequency at low shear. Relaxation time ($\lambda$) for PAM (0.2%) solution is 13s, whereas in presence of amine-coated PS (0.5%) $\lambda \sim 16$s.

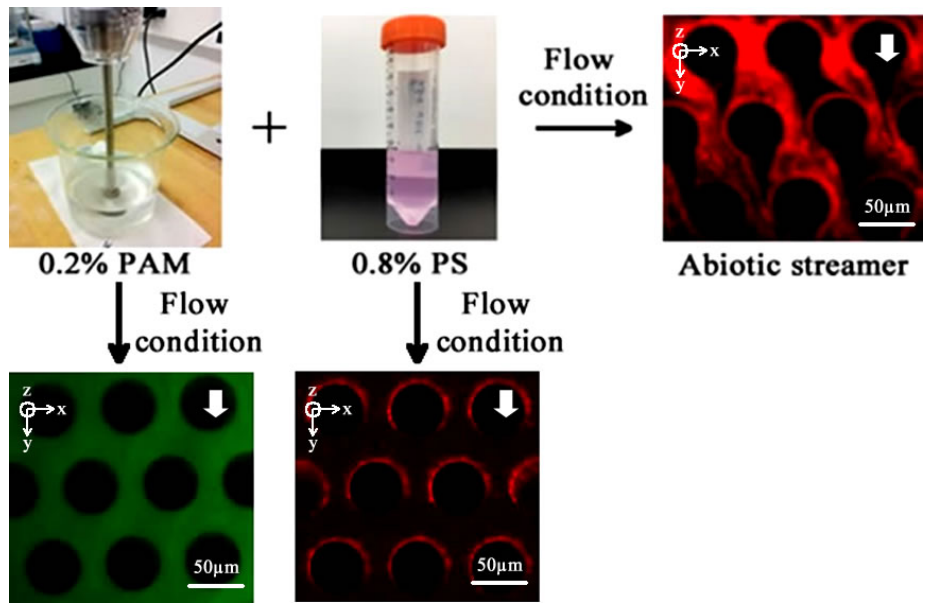

**Figure S3:** Control experiments for the investigation of streamer formation confirms that streamer forms for the combined flow of PAM (0.2% w/w) and PS (0.8%w/w) solution only.

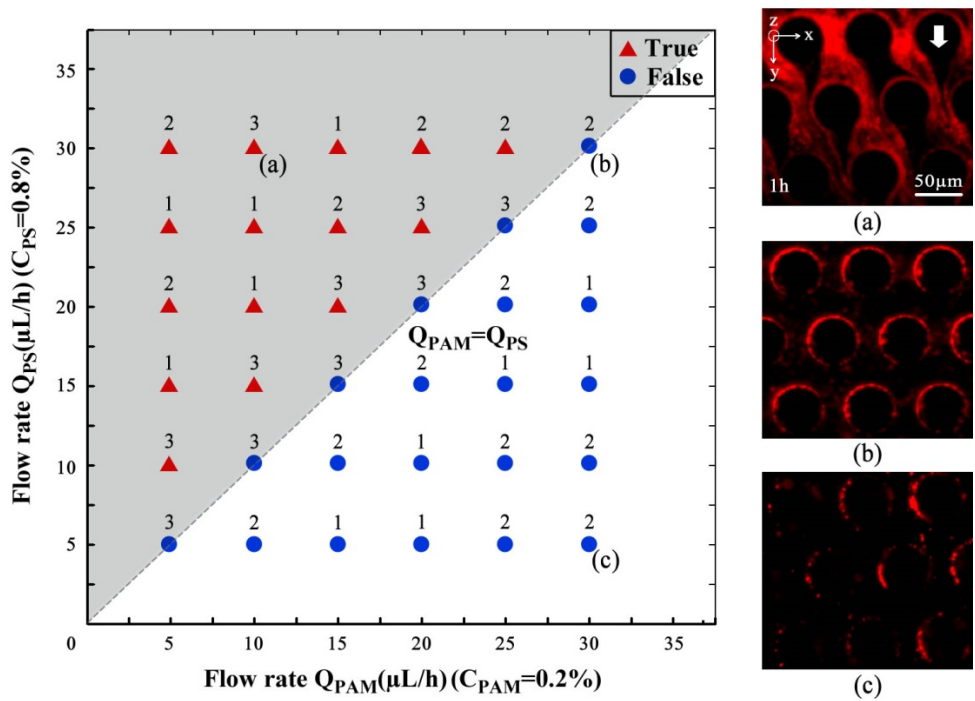

**Figure S4:** Repeat experiments for the phase diagram of streamer formation by varying flow rates when the concentration is fixed for $Q_{PAM}$ and $Q_{PS}$. Red triangle represents true streamer formation and blue circle indicates no streamer formation. Each number represents number of times the experiment was repeated. (a) Streamer formation after 1h for $Q_{PAM}$=10μL/h, $Q_{PS}$=30μL/h. (b) Transition phase for $Q_{PAM}$=$Q_{PS}$=30μL/h. (c) No streamer formation after 1h for $Q_{PAM}$=30μL/h, $Q_{PS}$=25μL/h. (d) Streamer formation (different morphology) after 1h for $Q_{PAM}$=5μL/h, $Q_{PS}$=20μL/h. (e) Transition phase for $Q_{PAM}$=$Q_{PS}$=15μL/h. (f) No streamer formation after 1h for $Q_{PAM}$=30μL/h, $Q_{PS}$=5μL/h.

## Supplementary video

**Supplementary video 1:** Video shows viscous abiotic streamer formation for short time scale and elastic abiotic streamer formation for long time scale. Video runs at real time and scale bars are 50μm.